\documentstyle[prl,aps,twocolumn,psfig]{revtex}
\begin{document}
\draft 
\title{Conductivity tensor of striped quantum Hall phases}
\author{Felix von Oppen$^1$, Bertrand I.\ Halperin$^2$, and Ady Stern$^{3}$}

\address{$^1$Institut f\"ur Theoretische Physik, Universit\"at zu K\"oln, 
 Z\"ulpicher Str.\ 77, 50937 K\"oln, Germany\\
$^2$Physics Department, Harvard University, Cambridge, 
Massachusetts, 02139\\
$^3$Department of Condensed Matter Physics, The Weizmann Institute of 
Science, 76100 Rehovot, Israel}

\date{\today}

\maketitle

\def\kf{k_{\mbox{\tiny{F}}}}              
\def\vf{v_{\mbox{\tiny{F}}}}
\def\rd{\rho_{\mbox{\tiny{D}}}}            
\def\cf{{\mbox{\tiny{CF}}}}
\def\saw{{\mbox{\tiny  SAW}}} 

\begin{abstract} 

  We study the transport properties of pinned striped quantum Hall
  phases. We show that under quite general assumptions, the
  macroscopic conductivity tensor satisfies a semicircle law. In
  particular, this result is valid for both smectic and nematic stripe
  phases, independent of the presence of topological and orientational
  defects such as dislocations and grain boundaries. As a special
  case, our results explain the experimental validity of a product
  rule for the dissipative part of the resistivity tensor, which was
  previously derived by MacDonald and Fisher for a perfect stripe
  structure.

\end{abstract}      
\bigskip       
\narrowtext


Recent experiments \cite{Lilly,Du} have revealed strikingly
anisotropic $dc$ transport properties of very clean two dimensional
(2D) electron systems when the Landau level filling factor is close to
$\nu=N+1/2$, and $N\ge 4$ is an integer.  It is believed that this is
related to previous theoretical proposals \cite{Fogler,Moessner} that
Coulomb interactions would lead to an instability towards
charge-density wave (CDW) formation in high, spin resolved, Landau
levels (LL).  Specifically, formation of a striped phase of the
uppermost Landau level was predicted when it is close to half filling,
while a ``bubble phase" should be favorable further away from half
filling.  The striped phase consists of one-dimensional stripes
alternating between the integer filling factors $N$ and $N+1$ with
period of order of the cyclotron radius $R_c$. In the bubble phase,
clusters of minority filling factor with size $R_c$ order in a
triangular lattice. These predictions, obtained within the
Hartree-Fock (HF) approximation, have also been supported by numerical
exact diagonalization studies \cite{Rezayi}.

At finite temperature and/or in the presence of disorder, the perfect
stripe ordering predicted by HF calculations will presumably
be destroyed.  The unidirectional CDW shares the symmetries of 2D
smectic liquid crystals \cite{Fradkin}. This implies that if there is
no external force tending to align the stripes, then a dislocation
will cost only a finite amount of energy and thus there will be a
finite density of dislocations at non-zero temperatures.  This is
expected to destroy translational long-range order, except at zero
temperature, but preserve quasi-long-range orientational order of the
remaining stripe segments, characteristic of a 2D nematic phase.  The
orientational order would be effectively locked in by any small added
anisotropy.  As the temperature becomes large enough, there will be a
Kosterlitz-Thouless transition to an isotropic state in which the
stripe segments lose their orientational order.  Short-range stripe
order should disappear completely only around the presumably much
higher HF transition temperature. 

Transport properties of the striped phases should be affected
by even small amounts of disorder on the substrate, which will pin the
stripe positions at low temperatures. Disorder should also lead to a
finite density of dislocations, even at zero temperature. Moreover,
since the forces aligning the stripes are believed to be very
weak, steps or other large-scale features of the GaAs-AlGaAs interface
may lead to large regions where the stripes are oriented 
differently from the average preferred direction.

In the present paper, we focus on the transport properties of general
striped quantum Hall phases, allowing in particular for the presence
of topological defects such as dislocations and grain boundaries.
Assuming that the defects are pinned by disorder, we find under quite
general assumptions (specified below) that the macroscopic
conductivity tensor $\hat\sigma^*$ satisfies the semicircle law
\begin{equation}
  \sigma^*_1\sigma^*_2+(\sigma^*_h-\sigma_h^0)^2 = (e^2/2h)^2,
\label{semicircle*}
\end{equation}
with $\sigma_h^0=(N+1/2)e^2/h$.  Here, we decomposed the macroscopic
conductivity $\hat\sigma^*=\hat\sigma^*_d+\sigma_h^*\hat\epsilon$ into
its dissipative part $\hat\sigma_d^*$ and Hall component
$\sigma_h^*\hat\epsilon$ with $\hat\epsilon$ the totally antisymmetric
tensor. The dissipative part is a real symmetric matrix with
eigenvalues $\sigma_1^*$ and $\sigma_2^*$.  Our derivation of
(\ref{semicircle*}) uses results obtained by Shimshoni and Auerbach
\cite{Shimshoni} for a model of a ``quantized Hall insulator''.
 
Following Refs.\ \cite{Fogler,Moessner} we assume that the system is
made up of regions (stripes or stripe segments) of filling factors $N$
and $N+1$, whose positions are fixed in space. The $N$ completely
filled Landau levels contribute to the conductivity tensor the pure
Hall response $N(e^2/h)\hat\epsilon$. The contribution of the
$(N+1)$th LL is due to the chiral edge modes at
the boundaries of each electron stripe (stripe of filling factor
$N+1$): transport occurs by motion along the edge modes and by
impurity scattering between them. We assume that scattering occurs
predominantly between {\it nearest-neighbor} edges, at rates
$1/\tau_e$ and $1/\tau_h$ across electron and hole stripes,
respectively. Recently, MacDonald and Fisher (MF) \cite{MacDonald}
have studied the transport properties of such a system, assuming a
uniform, topologically perfect stripe structure, with scattering rates
$\tau_e$ and $\tau_h$ that may depend on temperature, but are
independent of position.  Neglecting quantum interference effects and
taking the stripes along the $y$ direction, they find for the
conductivity tensor \cite{MacDonald}
\begin{eqnarray}
   \sigma_{xx}&=&
        {e^2\over h}{a\over v_F(\tau_e+\tau_h)}
      \nonumber\\
   \sigma_{yy}&=&
     ={e^2\over h}{v_F\over a}{\tau_e\tau_h\over\tau_e+\tau_h}
      \label{conductivity}\\
  \sigma_{xy}&=&{e^2\over h}\left(N+{\tau_e\over\tau_e+\tau_h}\right)
   \nonumber
\end{eqnarray}
with $a$ the period of the CDW and $v_F$ the velocity of propagation
along the edge channels. MF noticed \cite{MacDonald} that 
at the symmetric point, where $\tau_e=\tau_h$, this leads to the parameter-free
predictions
\begin{eqnarray}
    \sigma_{xx}\sigma_{yy}&=&(e^2/2h)^2
     \label{productrule} \\  
    \sigma_{xy}&=&{e^2\over h}(N+1/2),
\end{eqnarray}
independent of the period, Fermi velocity, or scattering rate.
If one assumes particle-hole symmetry in the partially full LL, then the symmetric
point will occur when the total filling factor is $N+1/2$. The
relation (\ref{productrule}) seems to be in reasonable agreement with the
experiment \cite{Eisenstein}.

In fact, the conductivity tensor (\ref{conductivity}) satisfies the more
general semicircle relation (\ref{semicircle*}), 
which is valid also away from the symmetric point.
In the homogeneous case discussed by MF, the macroscopic conductivity tensor is
identical to the local conductivity tensor, 
and we may write $\sigma^*_1=\sigma_{xx}$,
 $\sigma^*_2=\sigma_{yy}$,
and $\sigma^*_h=\sigma_{xy}$.

Since, as argued above, the real samples are likely to be quite far
from the perfect stripe structure, it may be surprising that the
product formula (\ref{productrule}) compares so favorably with
experiment. Moreover, we deduce from Eq.\ (\ref{conductivity}) that
for a perfect stripe structure the anisotropy at the symmetric point
is $\sigma_{yy}/\sigma_{xx}= (v_F\tau_e/a)^2$. The experimental
anisotropy in the conductivity is about five \cite{Lilly,Simon}, which
would imply that the electrons hop between edges after traveling only
a distance of a few cyclotron radii along the edge.  For such a
situation, quantum interference effects should be important,
particularly in view of the fact that the experiments are performed at
extremely low temperatures, which could lead to deviations from the
product rule (\ref{productrule}) and the semicircle law
(\ref{semicircle*}) \cite{footnote}.

In the following, we show in two different ways that the semicircle
and product rules embodied in (\ref{semicircle*}) are in fact valid
for much more general stripe structures. This makes the experimental
results consistent with a picture where electrons hop between edges
much more rarely, while the anisotropy is reduced by the presence of
defects such as dislocations and grain boundaries, which cause the
local orientation of the stripes to vary from one place to another. In
such a picture, neglecting quantum interference may indeed be
justified.

\begin{figure}

\centerline{\psfig{figure=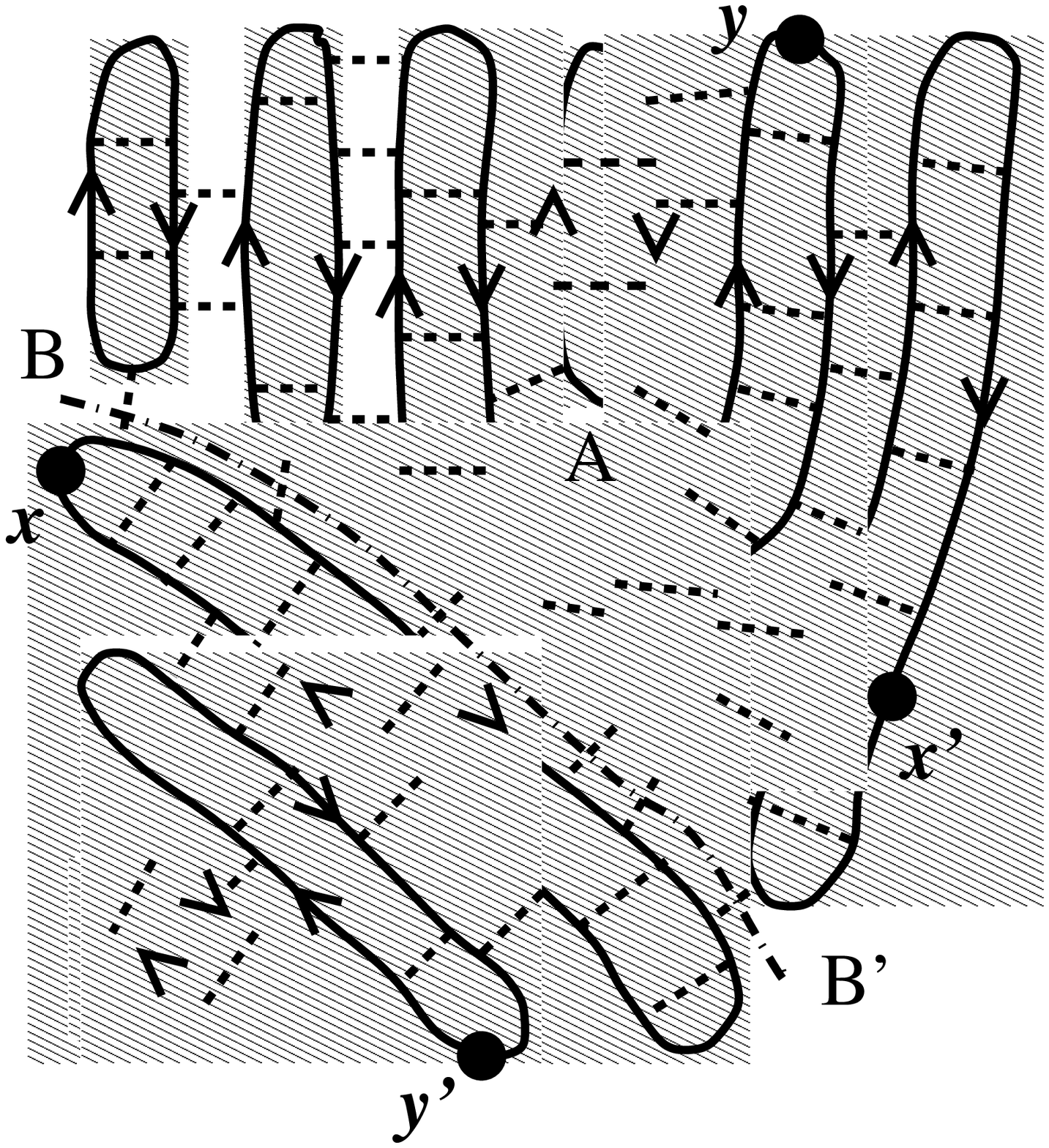,width=6.5cm,angle=270}}

\caption{Small sample of striped phase containing a dislocation (A) and a 
  large angle grain boundary (B-B$^\prime$).  Shaded and unshaded
  regions are incompressible strips with filling factors $N + 1$ and
  $N$ respectively; arrows show direction of electron flow on edge
  states, for positive magnetic field.  Dashed lines represent
  locations of scattering centers which cause electrons to tunnel
  between adjacent edge states.  Points $x$, $x^\prime$, $y$ and
  $y^\prime$ are contacts at the edge of the sample.}

\end{figure}

Consider a general stripe structure such as that shown in Fig.\ 1.  In
this figure, shaded and unshaded regions represent incompressible
states with filling factors $N+1$ and $N$, respectively.  The $N$
filled Landau levels are pure Hall conductors carrying current in
parallel to the $(N+1)$th LL. We can then focus on the role of the
uppermost LL by interpreting the shaded regions in the figure as an
incompressible state with $\nu=1$, while the unshaded regions are
insulators, with $\nu=0$. At the end, we add the Hall conductance
$Ne^2/h$ of the filled levels.

For the uppermost LL, we now consider a situation where a
current $I$ is passed between the contacts at $x$ and $x^\prime$
and the voltage $V$ is measured between contacts $y$ and $y^\prime$.
In general, $V$ has contributions from both the Hall and the
dissipative resistivities. 

The local current $i$ along a stripe-edge is related to the local
chemical potential $\mu$ (measured relative to the uniform equilibrium
chemical potential) by
\begin{equation}
   i= \mu \  {\rm sgn}(B) \  e^2/h .
\label{fundrel}
\end{equation}
Neglecting quantum interference effects, each scattering center between two adjacent electron stripes,
indicated by dashed lines in Fig.\ 1, can be characterized by a
scattering probability $p$.  The currents entering and leaving a
scattering center, cf.\ Fig.\ 2a, are related by current conservation
$i_1+i_2=i_3+i_4$ and by $i_3=(1-p)i_1+pi_2$ or $i_4=(1-p)i_2+pi_1$.
Equivalently, by using (\ref{fundrel}) we can characterize the scattering center by its
resistance $r=(h/e^2)(1-p)/p$, relating the voltage $v=\mu_2
-\mu_3 = \mu_4 - \mu_1 $ 
between edges to the current $j=i_3-i_1$ across.  In
terms of the quantities $j,v$ and $r$, stripe structures such as that
shown in Fig.\ 1 can be viewed as a type of classical resistor network subject
to Kirchhoff's laws and Eq.\ (\ref{fundrel}) relating currents and
voltages on the edges.

\begin{figure}

\centerline{\psfig{figure=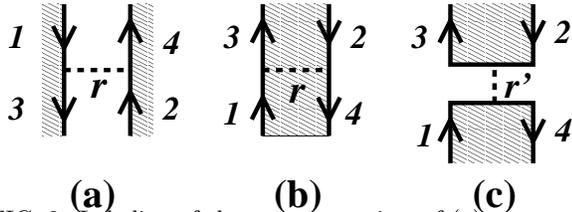,width=7.5cm,angle=270}}

\caption{Labeling of the current vertices of (a) a scattering center
  across a hole stripe and (b) across an electron stripe, for $B>0$. The
  resistances of the scattering centers are denoted by $r$. Any
  scattering center across an electron stripe as shown in (b) can be
  replaced by the equivalent dual representation shown in (c) in terms
  of a scattering center across a hole region with resistance
  $r^\prime$. The resistances of the two junctions are related by
  $r^\prime=(h/e^2)^2/r$.}

\end{figure}

We can ``eliminate" the scattering events across electron stripes from
the resistor network by using the following dual representation.  As
indicated in Fig.\ 2b and 2c, a scattering center across an electron stripe
described by $j,v,r$ is equivalent to a scattering center between
electron stripes described by $j^\prime,v^\prime,r^\prime$, where
$j^\prime=i_4-i_1$, $v^\prime=\mu_2 -\mu_4$, and
$r^\prime=v^\prime/j^\prime$. This is equivalent to the direct
relations $j^\prime=(e^2/h)v$, $v^\prime=(h/e^2)j$, and
$r^\prime=(h/e^2)^2/r$ between unprimed and primed parameters. The relation between $r$ and $r^\prime$ can be simply understood by noting that Fig. 2b maps onto Fig. 2c when $p$ is replaced by $1-p$. 

With this dual representation, the  network is defined by a set of
puddles (of filling factor $\nu=1$) each of which is encircled by one
chiral edge mode.  Electrons can scatter only between adjacent puddles as
described by resistors connecting the puddles. 
The network is planar in the sense that there are no
crossing resistor links between puddles.
The Hall voltage $V_H$ is even under simultaneous reversal of magnetic field $B$ and current $I$
so that it can be distinguished from the longitudinal voltage by the definition $V_H \equiv [V(B,I)+V(-B,-I)]/2$. By the definition of the network, the effect of changing the sign of $B$ is to reverse the directions of the arrows in the figure,
thus changing the sign in (\ref{fundrel}),
while leaving unchanged the values of the resistors $r$ and $r^\prime$.
This is consistent with the time-reversal properties of the
underlying microscopic Hamiltonian.

This resistor network (puddle model) has been studied in the context
of the quantized Hall insulator by Shimshoni and Auerbach
\cite{Shimshoni}. These authors prove that when quantum interference
between inter-puddle hopping events is neglected the Hall resistance
$R_H=V_H/I= h/e^2$.  Then, if the network is statistically
homogeneous, but not isotropic, and the sample is large compared to
any correlation length for fluctuations, we may adapt this result to
our problem and define macroscopic conductivity and resistivity
tensors for the uppermost LL, $\hat\sigma^u$ and $\hat\rho^u$.  Let us
choose the sample to have a Hall-bar geometry, aligned with the
principal axes of $\hat\rho^u$.  Then, using the Onsager symmetry
relations, we find
\begin{equation}
\rho^u_{yx}(B)= -\rho^u_{xy}(B) = -\rho^u_{yx}(-B)=h/e^2.
\label{rhoryx}
\end{equation}
This implies that det$(\hat\sigma^u) = \sigma^u_{xy} e^2/h$, and
the components of $\hat\sigma^u$ satisfy the semicircle law
(\ref{semicircle*}), with $\sigma^0_h = e^2/2h$.
Finally, if we add the parallel conductivity of the filled Landau
levels, the dissipative conductivity is unchanged, but the Hall
conductivity is shifted by $Ne^2/h$, leading to Eq.\ (\ref{semicircle*}).

We emphasize that the essential assumptions entering the proof are (a)
the neglect of quantum interference \cite{footnote}
and (b) the assumption that the
network is planar.
 The importance of the latter assumption
can already be understood in the context of the perfect stripe
structure. If we violate the assumption by including hopping to
next-nearest neighbor stripes, we find that this enhances the
conductivity in the $x$ direction while leaving the diffusion constant
in the $y$ direction unchanged. Thus, hopping to next-nearest neighbor stripes leads to
deviations from, e.g., the product rule (\ref{productrule}).
Also, the rule will be invalid if the temperature is so high that
dislocations are unpinned, and the stripe pattern itself can drift in
the presence of an applied electric field.  On the other hand, as in the
analysis of Shimshoni and Auerbach \cite{Shimshoni}, it is {\it not}
necessary to assume that the resistances $r$ and $r^\prime$ are ohmic,
i.e., independent of the magnitude of the current.  What is essential is
that the voltage across a resistor is reversed when the current and the
magnetic field are reversed.

It is possible to give an alternate (continuum) argument for the validity
of the semicircle law (\ref{semicircle*}) for $\hat\sigma^*$.  For
this argument, we assume that the most important role of defects is to
change both the local orientation of the CDW and the local scattering
rates between stripe edges. If the defect density is not too high,
these changes occur only on scales large compared to the cyclotron
radius and we can describe the system by a local conductivity tensor
$\hat\sigma({\bf r})$ whose principal axes and diagonal conductivities
are functions of position but which {\it locally} satisfies the semicircle
law:
\begin{equation}
  \sigma_1\sigma_2+(\sigma_h-\sigma_h^0)^2 = (e^2/2h)^2.
\label{semicircle}
\end{equation}
We now show, by an argument employing a duality
transformation introduced by Dykhne and Ruzin \cite{Dykhne}, that the
resulting macroscopic conductivity tensor satisfies the semicircle law
(\ref{semicircle*}).

The microscopic current distribution ${\bf J}({\bf r})$ is determined
by the equations
\begin{eqnarray}
   {\bf J}({\bf r})&=&\hat\sigma({\bf r}){\bf E}({\bf r}) \nonumber\\
   {\bf \nabla} \cdot {\bf J}({\bf r})&=&0   \\
    {\bf \nabla} \times{\bf E}({\bf r})&=&0.
    \nonumber
\end{eqnarray}
Then the dual system ${\bf J}^\prime,{\bf E}^\prime,\hat\sigma^\prime$
defined by the transformation
\begin{eqnarray}
    {\bf J}&=&a{\bf J}^\prime-b\hat\epsilon {\bf E}^\prime \\
    {\bf E}&=&c{\bf E}^\prime-d\hat\epsilon {\bf J}^\prime
\label{duality}
\end{eqnarray}
with $a,b,c,d$ arbitrary constants, satisfies the same set of equations
with $\hat\sigma^\prime=[a+d\hat\sigma\hat\epsilon]^{-1}(c\hat\sigma
+b\hat\epsilon)$. The same relation must hold for the corresponding 
macroscopic conductivities $\hat\sigma^*$ and $\hat\sigma^{*\prime}$,
because the macroscopic currents and voltages are just spatial averages of
the microscopic currents and voltages. 

Following Dykhne and Ruzin \cite{Dykhne}, we now choose the primed
system as the time reverse of the unprimed system,
$\sigma^\prime_{ij}=\sigma_{ji}$, and solve for $a,b,c,d$ in terms of
$\hat\sigma$.  Choosing $d=1$ without loss of generality, one finds
after some tedious but straight-forward algebra that 
$c=a$ and $b=\sigma_1\sigma_2+\sigma_h^2-2a\sigma_h$
with $a$ remaining arbitrary. This defines an allowed duality
transformation if $a,b,c,d$ are constants, independent of position.
Exploiting the semicircle law (\ref{semicircle}), we choose
$a=c=\sigma_h^0$, so that $b=(e^2/2h)^2-(\sigma_h^0)^2$,
independent of position. We can now combine the duality relation
between $\hat\sigma^*$ and $\hat\sigma^{*\prime}$ with the fact that
the two systems are related by time reversal,
$\sigma^{*\prime}_{ij}=\sigma^*_{ji}$, and find that the macroscopic
conductivity tensor $\hat\sigma^*$ satisfies the semicircle law
(\ref{semicircle*}).

It is an interesting question whether the experimental validity of the
product rule is specific to the CDW model involving local stripe
ordering with period comparable to the cyclotron radius. 
For example, one might hypothesize that close to half filling, the
system phase-separates, on a scale much larger than the cyclotron
radius, into quantized Hall regions with $\nu=N$ and $\nu=N+1$, with
only thin boundaries between the regions.  If the shapes of the regions
are elongated, with a preference for one particular direction in space,
one would find an anisotropic macroscopic conductivity tensor which
would obey the semicircle law (\ref{semicircle*}) under appropriate
conditions.  In such a model, however, one obtains very small values
of  $\sigma^*_1$ and $\sigma^*_2$, except very close to the percolation
threshold for the two phases.  Thus one would expect the transition 
between Hall plateaus corresponding to $\nu=N$ and $\nu=N+1$ to occur in a very
narrow interval of magnetic field, which is contrary to experimental
observations in the samples of interest \cite{Lilly,Du}.

In order to interpret experimental results in terms of an effective
macroscopic conductivity tensor $\hat\sigma^*$, which is anisotropic but
spatially uniform, it is important that there be reasonable
equilibration between the edge states of the filled LL and the electrons
of the partially filled LL, at all points on the sample boundary.  
If the density profile at the sample edge is sufficiently
gradual, however, the spatial separation of the edge states may cause
this equilibration to fail at low temperatures.  The use of an effective
conductivity tensor can be checked, in principle, by comparing samples
with different aspect ratios or different contact locations.

The theory presented here does not address a number of issues raised
by the experimental observations \cite{Lilly,Du}.  The mechanism which
causes the stripes to line up preferentially with a particular axis of
the GaAs substrate is not well understood.  There is currently no
explanation for the observation that the resistance anisotropy has a
prominent dependence on whether the Fermi energy is in the lower
or upper spin component of a LL.  An explanation for the observed
non-linearity in the resistivities has been proposed by MF
\cite{MacDonald}, involving quantum fluctuations of the
Luttinger-liquid type; however, there are not yet detailed predictions
for the full temperature and current dependences which might be
compared with experiment.

In summary, we have studied the transport properties of general
striped quantum Hall phases. Assuming that the stripes are pinned by
disorder and that scattering between stripes can be considered
classically, we have established rather generally the validity of a
semicircle law (\ref{semicircle*}) for the macroscopic conductivity
tensor. Our results provide an explanation for the experimental
validity of the product rule (\ref{productrule}) which is a special
case of (\ref{semicircle*}).  It would be very interesting to check if
the more general semicircle law (\ref{semicircle*}) is also obeyed in
experiment.

We are grateful to J. Eisenstein, B. Huckestein, M. Shayegan, S.
Simon, D. Shahar and H. Stormer for helpful discussions.  The work was
supported in part by SFB 341, NSF grant DMR-94-16910, US-Israel BSF
grant 98-354, DIP-BMBF grant and the Israeli Academy of Science.
Initial phases of the work were carried out at the ITP, Santa Barbara,
in August 1998, with support from NSF grant PHY94-07194.


\begin{references}

\bibitem{Lilly} M.P.\ Lilly, K.B.\ Cooper, J.P.\ Eisenstein, L.N.\ Pfeiffer, 
and K.W.\ West, Phys.\ Rev.\ Lett.\ {\bf 82}, 394 (1999).

\bibitem{Du} R.R.\ Du, D.C.\ Tsui, H.L.\ Stormer, L.N.\ Pfeiffer,
K.W.\ Baldwin, and K.W.\ West, Solid State Comm.\ {\bf 109}, 389 (1999).

\bibitem{Fogler} M.M.\ Fogler, A.A.\ Koulakov, and B.I.\ Shklovskii,
Phys.\ Rev.\ B {\bf 54}, 1853 (1996); A.A.\ Koulakov, M.M.\ Fogler,
and B.I.\ Shklovskii, Phys.\ Rev.\ Lett.\ {\bf 76}, 499 (1996). 

\bibitem{Moessner} R.\ Moessner, J.T.\ Chalker, Phys.\ Rev.\ B
{\bf 54}, 5006 (1996). 

\bibitem{Rezayi} E.H.\ Rezayi, F.D.M.\ Haldane, and K.\ Yang, 
Phys.\ Rev.\ Lett.\ {\bf 83}, 1219 (1999).

\bibitem{Fradkin} E.\ Fradkin, S.A.\ Kivelson, Phys.\ Rev.\ B
{\bf 59}, 8065 (1999).

\bibitem{Shimshoni} E.\ Shimshoni and A.\ Auerbach, Phys.\ Rev.\ B {\bf 55},
9817 (1997).


\bibitem{MacDonald} A.H.\ MacDonald and M.P.A.\ Fisher, cond-mat/9907278.

\bibitem{Eisenstein} J.\ Eisenstein, private communication.

\bibitem{Simon} S.H.\ Simon, cond-mat/9903086.

\bibitem{footnote} In general, the semicircle law does not apply to a
  quantum mechanical model, where quantum localization or interference
  effects may be important. [See L.P.\ Pryadko and A.\ Auerbach,
  Phys.\ Rev.\ Lett.\ {\bf 82}, 1253 (1999)].  However, the size of
  the deviation is highly model-dependent, and could be small, so one
  cannot necessarily exclude quantum interference effects on the basis
  of the experimental results.


\bibitem{Dykhne} A.M.\ Dykhne, I.M.\ Ruzin, Phys.\ Rev.\ B {\bf 50},
2369 (1994).

\end{references}
\end{document}